\documentclass{emulateapj}

\usepackage{graphics}
\usepackage{epsfig}
\usepackage{natbib}
\shorttitle{SLICE Observations of H$_{2}$ Excitation}
\shortauthors{France et al.}
\begin{document}

\title{H$_{2}$ Excitation Structure on the 
Sightlines to $\delta$ Scorpius and $\zeta$ Ophiucus~--~First Results from the Sub-orbital Local Interstellar Cloud Experiment}


\author{
Kevin France\altaffilmark{1,4},
Nicholas Nell\altaffilmark{1}, 
Robert Kane\altaffilmark{1}, 
Eric B. Burgh\altaffilmark{2},
Matthew Beasley\altaffilmark{3}
and James C. Green\altaffilmark{1}
}


\altaffiltext{1}{Center for Astrophysics and Space Astronomy, University of Colorado, 389 UCB, Boulder, CO 80309; kevin.france@colorado.edu}
\altaffiltext{2}{SOFIA/USRA, NASA Ames Research Center, M/S N232-12, Moffett Field, CA 94035, USA}
\altaffiltext{3}{Planetary Resources, Inc., 93 S Jackson St \#50680, Seattle, Washington 98104-2818}
\altaffiltext{4}{NASA Nancy Grace Roman Fellow}	


\begin{abstract}

We present the first science results from the Sub-orbital Local Interstellar Cloud Experiment (SLICE): moderate resolution 1020~--~1070~\AA\ spectroscopy of four sightlines through the local interstellar medium.   High signal-to-noise (S/N) spectra of $\eta$~Uma, $\alpha$~Vir, $\delta$~Sco, and $\zeta$~Oph were obtained during a 21 April 2013 rocket flight.  
The SLICE observations constrain the density, molecular photoexcitation rates, and physical conditions present in the interstellar material towards $\delta$~Sco and $\zeta$ Oph.   Our spectra indicate a factor of two lower total N(H$_{2}$) than previously reported for $\delta$~Sco, which we attribute to higher S/N and better scattered light control in the new SLICE observations. 
We find N(H$_{2}$)~=~1.5~$\times$~10$^{19}$ cm$^{-2}$ on the $\delta$ Sco sightline, with kinetic and excitation temperatures of 67 and 529~K, respectively, and a cloud density of $n_{H}$~=~56 cm$^{-3}$.
Our observations of the bulk of the molecular sightline toward $\zeta$ Oph are consistent with previous measurements (N(H$_{2}$)~$\approx$~3~$\times$~10$^{20}$ cm$^{-2}$ at $T_{01}$(H$_{2}$)~=~66 K and $T_{exc}$~=~350~K).  However, we detect significantly more rotationally excited H$_{2}$ towards $\zeta$~Oph than previously observed.  We infer a cloud density in the rotationally excited component of $n_{H}$~$\approx$~7600 cm$^{-3}$ and suggest that the increased column densities of excited H$_{2}$ are a result of the ongoing interaction between $\zeta$~Oph and its environment; also manifest as the prominent mid-IR bowshock observed by $WISE$ and the presence of vibrationally-excited H$_{2}$ molecules observed by $HST$.    

\end{abstract}

\keywords{instrumentation: spectrographs --- ISM: lines and bands --- ISM: molecules --- stars: individual ($\eta$~Uma (HD 120315), $\alpha$~Vir (HD 116658), $\delta$~Sco (HD 143275),  $\zeta$~Oph (HD 149757))}
\clearpage

\section{Introduction}

Molecular hydrogen (H$_{2}$) is ubiquitous in space, comprising the majority of the mass of the interstellar medium, protostellar/protoplanetary disks, and giant planets.  
The characteristics of the molecular phases of diffuse and translucent interstellar clouds have largely been determined by far-UV observations of H$_{2}$ absorption lines on the sightlines to hot background stars. The  Lyman and Werner band systems (observed primarily at 912~--~1120~\AA) have been widely studied by sounding rockets~\citep{carruthers70,jenkins89,france04}, $Copernicus$~\citep{spitzer74b,savage77}, the shuttle-borne IMAPS, HUT, and ORFEUS instruments~\citep{jenkins97,bowers95,lee02}, and most recently the {\it Far-Ultraviolet Spectroscopic Explorer} ($FUSE$; Rachford et al. 2002, 2009; Burgh et al. 2007).\nocite{rachford02,rachford09,burgh07}  
Since the first high-resolution studies with $Copernicus$ in the 1970s, it has been known that many interstellar sightlines display a multi-component H$_{2}$ population structure~\citep{spitzer73}.  The lowest rotational levels ($J^{''}$~=~0 and 1) are collisionally populated and representative of the kinetic temperature of the cloud, while the intermediate rotational levels ($J^{''}$~$\approx$~2~--~7) follow a higher temperature, possibly non-thermal distribution.  The H$_{2}$ population structure is a powerful diagnostic for understanding the physical conditions of an interstellar cloud, providing constraints on the local UV radiation field, the density, and rate of H$_{2}$ formation on grains.  

Despite the extensive study of the characteristics of interstellar H$_{2}$, there is still considerable uncertainty about the physical mechanism responsible for the intermediate-$J$ excitation observed in molecular clouds.   The canonical view for the excitation of the $J^{''}$~=~2~--~7 levels is that UV photons, either from the UV-luminous target star or an enhancement in the local interstellar radiation field, fluorescently excite these states and they are observed in absorption owing to the very long radiative decay times from these levels~\citep{spitzer74,vdb86,browning03}.  Excess energy associated with H$_{2}$ formation has also been discussed by several authors as the source of the intermediate-$J$ population~\citep{jura74,lacour05}.  There is growing evidence that fluorescent excitation and grain formation energy alone may be insufficient to reproduce the large column densities observed in these levels~\citep{gry02,sonnentrucker03}.  Interestingly, deep  $Spitzer$-IRS observations~\citep{ingalls11} of pure rotational emission ($\Delta$$J$~=~4~$\rightarrow$~2), ($\Delta$$J$~=~3~$\rightarrow$~1), and ($\Delta$$J$~=~2~$\rightarrow$~0) have also found that UV-pumping cannot reproduce the observed column densities of the $J^{''}$~=~2~--~4 levels.   These authors have argued that collisional processes such as the recent passage of a supernova blast wave or the dissipation of interstellar turbulence is the more likely mechanism by which these states are populated.  

The nearest O and B stars 
have not been observed by modern UV space instruments employing low-scatter gratings, large instrumental bandpasses (for optimal continuum normalization), and large-format low-background detectors.  
The stringent bright-object limits of $FUSE$ prevented the observation of most OB-stars in the local interstellar medium (LISM; $d$~$\lesssim$~200~pc).  The advantages of studying H$_{2}$ excitation in the LISM include $a)$ the line-of-sight velocity structure is simple; there is typically only one molecular cloud between the edge of the local bubble 
the target star, $b)$ the average interstellar radiation field is relatively well-constrained in the LISM, and $c)$ kinematic structures in the LISM can be spatially resolved, allowing one to identify potential interactions between hot stars and the ambient ISM.  Most observations of H$_{2}$ in the LISM date from $Copernicus$, which recorded spectra in roughly~1~\AA\ scans with large and variable backgrounds from charged particles and scattered light.
In order to obtain simultaneous observations of multiple-H$_{2}$ absorption lines and high-fidelity continuum normalization for local ISM targets, we have developed and launched the Sub-orbital Local Interstellar Cloud Experiment (SLICE).   In this Letter, we present the first flight results from SLICE: new measurements and upper limits on H$_{2}$ column densities for four hot stars at $d$~$\leq$~150 pc ($\eta$~Uma, $\alpha$~Vir, $\delta$~Sco, and $\zeta$~Oph) and new determinations of the H$_{2}$ populations on the sightlines to $\delta$~Sco and $\zeta$~Oph.

\section{SLICE Instrument and Observations}

SLICE is a rocket-borne instrument consisting of a far-UV optimized telescope, a spectrograph, and an electronics package that interfaces with a NASA telemetry system to downlink the spectroscopic data in real-time (Figure 1).  The telescope is a 20-cm primary diameter Cassegrain.  The telescope and spectrograph are matched systems at $f$/7.  The aluminum telescope and grating optics employ LiF-overcoatings to maximize sensitivity in the 1020~--~1070~\AA\ SLICE bandpass.    Stellar spectra are focused onto a spectrograph entrance slit cut into a mirrored slit-jaw, which is imaged by an aspect camera to assist in target acquisition during the flight.  The dispersing instrument is a modified Rowland circle spectrograph, which feeds a `chevron/z-stack' microchannel plate intensified photon-counting detector.  The SLICE payload achieves a detector resolution-limited resolving power of $R$~=~5300 ($\Delta$$\lambda$~=~0.2~\AA, $\Delta$$v$~$\sim$~60 km s$^{-1}$) across the bandpass, with a total system effective area, $A_{eff}$~$\approx$~2 cm$^{-2}$.   

\begin{figure}
\begin{center}
\epsfig{figure=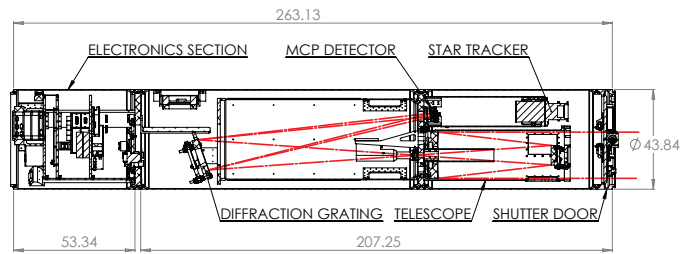,width=3.6in,angle=00}
\vspace{+0.0in}
\caption{
\label{cosovly} Schematic view of the Sub-orbital Local Interstellar Cloud Experiment (SLICE).  Dimensions are in centimeters.  
 }
\end{center}
\end{figure}

\begin{figure*}
\begin{center}
\epsfig{figure=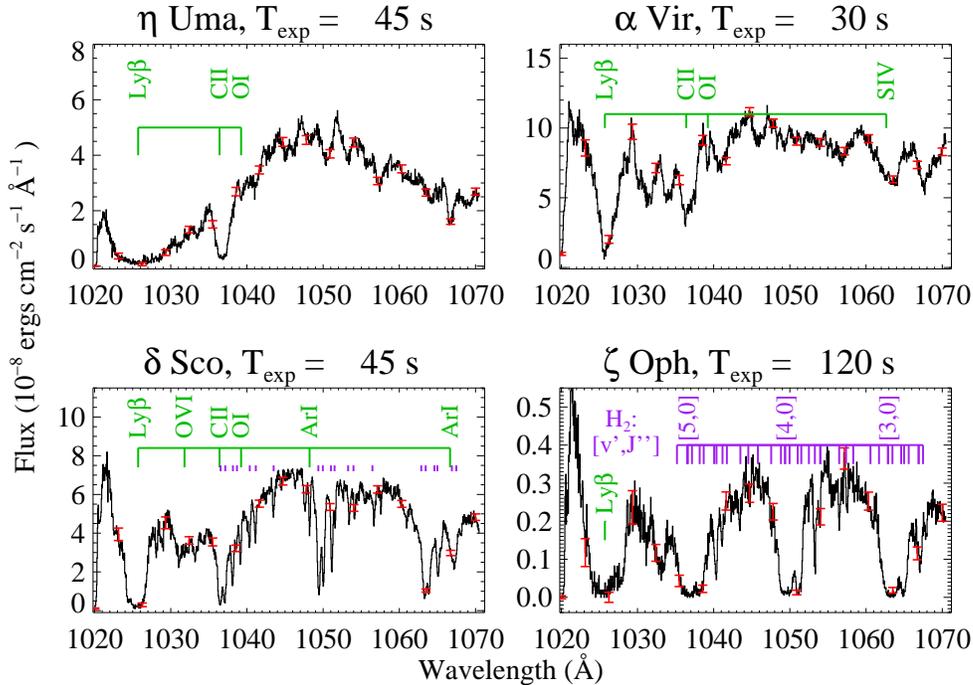,width=3.6in,angle=90}
\vspace{+0.0in}
\caption{
\label{cosovly} Flux calibrated 1020~--~1070~\AA\ spectra of the SLICE targets.  Representative error bars are shown in red and prominent stellar and interstellar lines are labeled.
 }
\end{center}
\end{figure*}

SLICE was launched aboard a Terrier-Black Brant IX sounding rocket from White Sands Missile Range at 02:00 MDT on 21 April 2013 as part of NASA mission 36.271~UG.    The detector acquired data for $\approx$~400s during the flight, with acquisitions divided into ``on-target'' times when the stars were in the spectrograph aperture and the pointing was stable.  This resulted in exposure times of $T^{i}_{exp}$~=~45s, 30s, 45s, and 120s for $i$ = $\eta$~Uma, $\alpha$~Vir, $\delta$~Sco, and $\zeta$~Oph, respectively (see Table 1).  SLICE achieved signal-to-noise ratios (S/Ns) of $>$~30 per pixel on the first three targets and S/N~$\approx$~12 per pixel over the middle of the SLICE bandpass (1045~--~1060~\AA) on $\zeta$~Oph.   The SLICE wavelength solution was established pre- and post-flight 
using collisionally-excited H$_{2}$ emission spectra as a reference.   The wavelength solution has an accuracy of approximately 2 pixels across the bandpass ($\approx$~0.1~\AA), and a zero-point offset was applied to each target to account for the exact location of the star in the spectrograph aperture.  The data were downlinked through the telemetry system as an [$x$,$y$,$t$] photon list. 

\section{Analysis} 

\subsection{Spectral Extraction and Flux Calibration}

An [$x_{i}$,$y_{i}$] photon list for each star was created by isolating the ``on-target'' times $\Delta$$t_{i}$ = $T^{i}_{o}$ + $T^{i}_{exp}$, where 
$T_{o}$ is the start time for each integration.   Each [$x_{i}$,$y_{i}$] photon list was collapsed over the cross-dispersion astigmatic height ($\approx$~3.5 mm) and divided by  $T^{i}_{exp}$ to produce one-dimensional stellar spectra in units of [counts s$^{-1}$].   
We combined our highest S/N observation with stellar models to create a flux calibration curve for the SLICE instrument.  
We modeled the $IUE$ spectra of $\alpha$ Vir (Spica) with the TLUSTY B-star models~\citep{lanz07}, finding good agreement with the observed 1150~--~1250~\AA\ data for a model with $T_{eff}$~=~24,000 K, log($g$)~=~3.33,  $Z$~=~0.5$Z_{\odot}$, and $v_{turb}$~=~2 km s$^{-1}$, scaled to an 1150~\AA\ flux, $F$(1150)~=~8.3~$\times$~10$^{-8}$ erg cm$^{-2}$ s$^{-1}$ \AA$^{-1}$.   A spline function was fitted to the ratio of stellar model-to-SLICE observations for 15 points relatively unobscured by strong stellar or interstellar absorption lines, creating a smooth flux calibration curve from 1020~--~1070~\AA, in units of [erg cm$^{-2}$ s$^{-1}$ \AA$^{-1}$ / counts s$^{-1}$].   The flux-calibrated spectra of $\eta$~Uma, $\alpha$~Vir, $\delta$~Sco, and $\zeta$~Oph are shown in Figure 2. 

\subsection{H$_{2}$ Profile Fitting and Excitation Diagrams}

{\it Column Densities~--~} The two primary means of determining column densities for individual H$_{2}$ rotational levels from absorption line spectroscopy are curve-of-growth fitting and profile fitting.  While in principle we have coverage of most of four ($v^{'}$~--~0) Lyman bands ($v^{'}$~=~6, 5, 4, 3; with R(0) wavelengths 1024.37~\AA, 1036.54~\AA, 1049.37~\AA, and 1062.88~\AA, respectively) that should permit a robust curve-of-growth analysis, $v^{'}$~=~6 is contaminated with stellar+interstellar \ion{H}{1} Ly$\beta$, $v^{'}$~=~5 is contaminated with interstellar \ion{C}{2} $\lambda$$\lambda$1036, 1037, and $v^{'}$~=~3 is contaminated by stellar lines and interstellar \ion{Ar}{1} $\lambda$1066.  At the spectral resolution of SLICE, only the (4~--~0) band is ``clean'' for H$_{2}$ line measurements of low-to-intermediate rotational levels ($J^{''}$~=~0~--~7); therefore, we employed spectral profile fitting to measure the H$_{2}$ column density distributions on the lines-of-sight to our target stars.  

We constructed a multi-component H$_{2}$ absorption line fitting routine combining the $H_{2}ools$ optical depth templates~\citep{mccandliss03} and the MPFIT least-squares minimization routines in IDL~\citep{markwardt09}.  Our method takes the theoretical line shape of each H$_{2}$ rotational level for a given column density (N(H$_{2}$[0,$J^{''}$])~$\equiv$~N(J${''}$)) and Doppler $b$-value, convolves the synthetic spectrum with the line-spread-function of the SLICE instrument and simultaneously varies all parameters until a best-fit value is found.   A single molecular component is known to dominate each sightline~\citep{spitzer74b,morton75}, thus a single component was employed for our fitting.  For computational simplicity, we restrict our fits to a single $b$-value for all of the H$_{2}$.  To constrain the $b$-value, we ran H$_{2}$ model fits over a portion of the (4~--~0) band that restricted the fits to $J^{''}$~=~2~--~6, thereby constraining the $b$-value for the higher rotational levels most sensitive to changes in $b$.   Our fits for $\delta$~Sco and $\zeta$~Oph found best-fit $b$ between 3 and 4 km s$^{-1}$, consistent with typical $b$ values for H$_{2}$ in the local ISM~\citep{lehner03} and previous curve-of-growth measurements of H$_{2}$ on these sightlines~\citep{spitzer74b,snow88}; therefore we present the H$_{2}$ results for a best-fit $b$-value of 4 km s$^{-1}$ for both stars.  The single $b$-value approximation may introduce systematic error in our N($J^{''}$) determination because it has been shown that on some sightlines, the higher rotational levels ($J^{''}$~$\approx$~4~--~6) are better fit with $b$-values a factor of $\sim$~2 larger than for $J^{''}$~=~2 (see, e.g., Lacour et al. 2005 and references therein).   The determination of the total column density is not affected because the low-$J^{''}$ lines that dominate the total column are highly damped and insensitive to $b$.   
We estimate an uncertainty in the continuum placement of $\pm$~3\% for the (4~--~0) band, which introduces error less than the fitting uncertainty for $J^{''}$~=~0~--~2.  Underestimating the continuum level can have a large impact on N($J^{''}$~=~3~--~5), where uncertainties $\sim$~1 dex can be introduced.   

\begin{figure}
\begin{center}
\epsfig{figure=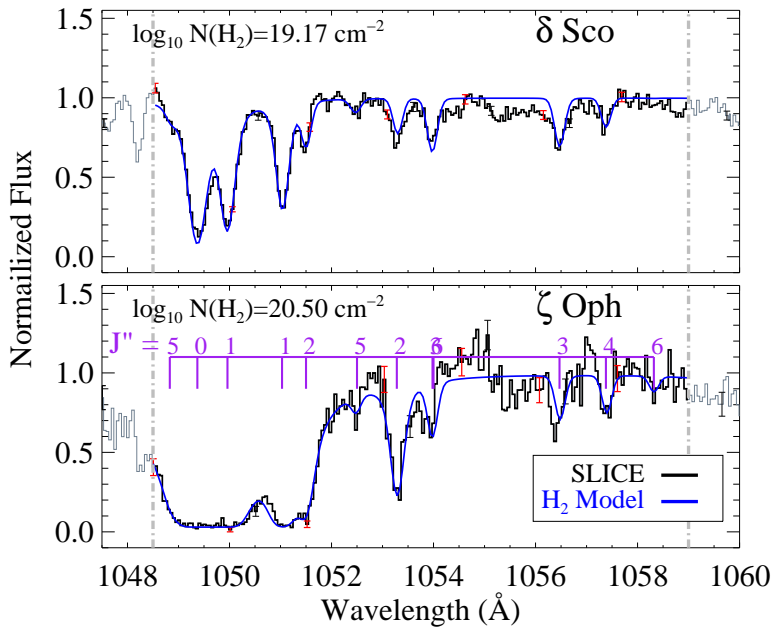,width=2.88in,angle=90}
\vspace{+0.5in}
\caption{
\label{cosovly} Synthetic H$_{2}$ (4~--~0) band spectral fits for $\delta$~Sco and $\zeta$~Oph.  
The best-fit Doppler velocity for these fits is~$b$~$\approx$~4 km s$^{-1}$.  
}
\end{center}
\end{figure}


We present the total H$_{2}$ column densities in Table 1.  For the two targets inside the local bubble ($\eta$ Uma and $\alpha$ Vir), no H$_{2}$ was observed.  
The upper limits for these sightlines are log$_{10}$ N(H$_{2}$) $<$~14.86 and $<$~14.67, respectively.   
The normalized spectra and best-fit models for $\delta$~Sco and $\zeta$~Oph are shown in Figure 3.  We measure total columns for these sightlines log$_{10}$ N(H$_{2}$)~=~19.17~$\pm$~0.06  and 20.50~$\pm$~0.12, respectively.  Taking the neutral hydrogen columns from~\citet{diplas94}, we compute molecular fractions,
\begin{equation}
 f(H_{2})~=~\frac{2N(H_{2})}{N(HI) + 2N(H_{2}) } 
\end{equation}
of $f$(H$_{2}$)~=~0.026~$\pm$~0.006 and $f$(H$_{2}$)~=~0.56~$\pm$~0.20 for the $\delta$ Sco and $\zeta$ Oph sightlines.  

{\it Kinetic and Excitation Temperatures ~--~}  
Radiative transitions from $J^{''}$~=1~$\rightarrow$~ $J^{'}$~=~0 are forbidden and the quadrupole transitions $J^{''}$~=~2~$\rightarrow$~$J^{'}$~=~0
are very slow ($A_{2 \rightarrow 0}$~$\approx$~3~$\times$~10$^{-11}$~s$^{-1}$; Wolniewicz et al. 1998); at the densities of these clouds (10$^{1}$~--~10$^{4}$ cm$^{-3}$), collisions control the level populations of the $J^{''}$ = 0, 1, and (sometimes) 2 states.\nocite{wolniewicz98}  
The kinetic temperature, $T_{01}$, of the cloud can therefore be determined from the ratio of the column densities in these levels, 
\begin{equation}
N(J^{''}=1)/N(J^{''}=0) = \frac{g_{1}}{g_{0}} e^{(-E_{01}/kT_{01})} = 9 e^{(-171 K/T_{01})}
\end{equation}  
where $g_{J''}$ is the statistical weight.  Following this prescription, we calculate $T_{01}$ kinetic temperatures on the $\delta$~Sco and $\zeta$~Oph sightlines of 67~$\pm$~1 K and 66~$\pm$~3 K, respectively.  

We characterize the higher rotational levels by creating H$_{2}$ excitation diagrams, as shown in Figure 4.  
The higher lying levels can be fit with an ``excitation temperature'' ($T_{exc}$) that describes the rotational excitation of the higher-$J$ states. 
The slope of the excitation diagram for the intermediate rotational states ($J^{''}$ = 4~--~6 for $\delta$~Sco and $J^{''}$ = 3~--~6 for $\zeta$~Oph) can be directly related to $T_{exc}$.  A least-squares linear fitting routine was used to determine $T_{exc}$ for each sightline, finding $T_{exc}$ = 529~$\pm$~99 K and $T_{exc}$ = 350~$\pm$~75 K for $\delta$~Sco and $\zeta$~Oph, respectively. 


\section{Discussion}

\subsection{Comparison with Previous Observations } 

The H$_{2}$ properties derived from the SLICE observations are in rough agreement (factor of 2) with most previous observations of the molecular sightlines toward $\delta$ Sco and $\zeta$~Oph; however, there are some notable differences:  

{\it $\delta$~Sco~--~}The total H$_{2}$ column density derived from SLICE is approximately a factor of two lower than obtained from $Copernicus$ and previous suborbital observations~\citep{savage77,snow88}, while the cloud kinetic temperatures ($T_{01}$) and Doppler $b$-values are comparable.  Conversely, we find a roughly factor of 2~--~3 larger $J^{''}$~$\geq$~4 excitation temperature (529 K) for the $\delta$~Sco sightline than found by either Spitzer et al. (1974; 318 K) or Snow et al. (1998; 210 K).  
The largest sources of uncertainty for the measurement of H$_{2}$ column densities are the continuum normalization and the S/N of the data.  The $Copernicus$ measurements were made from narrow-bandpass spectral scans that compromise a robust continuum determination, and the sounding rocket data presented by Snow et al. were of low S/N and affected by a poorly-characterized source of instrumental scattered light, which made continuum placement challenging.  Therefore, we believe the H$_{2}$ excitation results derived from the SLICE data are more accurate.  

{\it $\zeta$~Oph~--~}We found a somewhat ($\sim$~30~\%) lower N(H$_{2}$) than $Copernicus$~\citep{spitzer73,savage77}, although this may be attributable to line-blending between the \ion{Ar}{1} $\lambda$~1048~line and the heavily damped (4~--~0) R(0) line in the lower resolution SLICE data.  As a check, we isolated the fits to the (3~--~0) R(0), R(1), and P(1) lines (recall that stellar and interstellar spectral contamination prevents measurement of the entire (3~--~0) Lyman band).  The (3~--~0) $J^{''}$~=~0 and 1 fits found column densities approximately 30\% larger than the $Copernicus$ value.  Therefore, the average SLICE N(H$_{2}$) is consistent with the average $Copernicus$ result.  
  We find a nearly identical excitation temperature ($T_{exc}$~=~324 vs. 350 K), although the SLICE observations yield N($J^{''}$) values 0.3~--~1.0 dex higher for $J^{''}$~=~2~--~6.  This result is surprising given that we find a nearly identical $b$-value for these lines ($b$~=~3.8 vs. 4 km s$^{-1}$).  
  We tested this directly by comparing the $J^{''}$~=~2~--~4 equivalent widths measured from the SLICE data with those given in~\citet{spitzer74b}.  We find equivalent widths~$\sim$~15~--~25\% larger in the SLICE data, which is approximately the expected increase in equivalent width (4~--~37\%) resulting from the larger SLICE column densities.  Therefore, we conclude that the enhanced columns of rotationally excited H$_{2}$ are a physical effect.  

\begin{figure*}
\begin{center}
\epsfig{figure=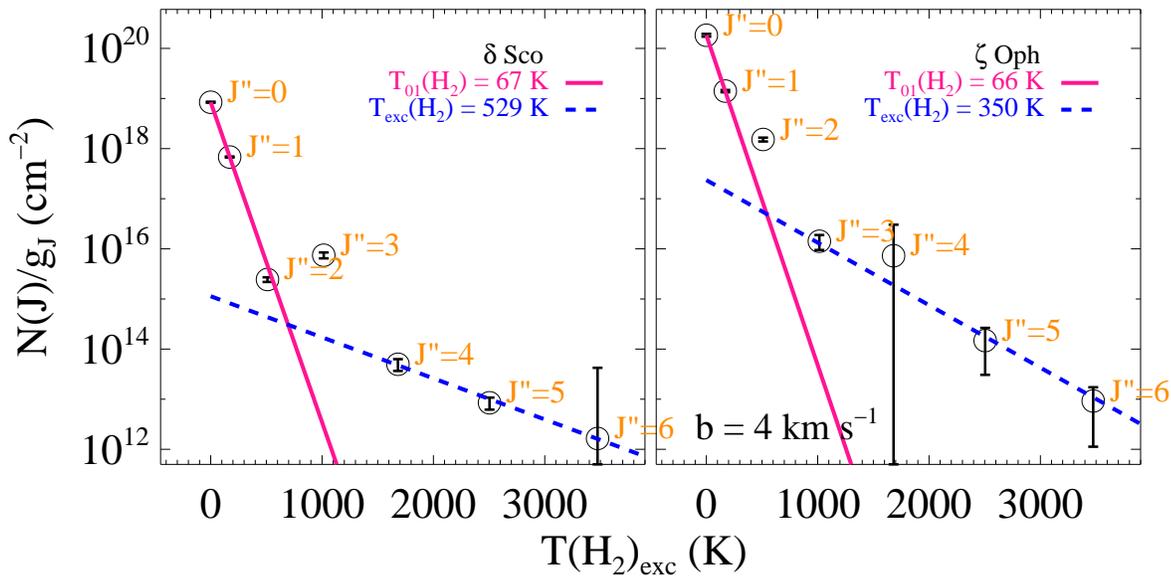,width=3.5in,angle=90}
\vspace{+0.2in}
\caption{
\label{cosovly} 
H$_{2}$ Excitation diagrams for  $\delta$~Sco and $\zeta$~Oph demonstrate the two-temperature population structure of these molecular clouds.  
 }
\end{center}
\end{figure*}

\subsection{Physical Cloud Conditions}	

Combining the N($J^{''}$) measurements derived from the SLICE observations with a plane-parallel interstellar cloud model, we are able to estimate the physical properties of the clouds on the $\delta$~Sco and $\zeta$~Oph sightlines.  Under the assumption of constant density, the product of the H$_{2}$ formation rate on grains ($R_{form}$) and the total particle density ($n_{H}$), $R_{form}$$n_{H}$, can be related to the ratio of H$_{2}$ column density in the $J^{''}$~=~4 level to the atomic hydrogen column density~\citep{jura75b}, and can be rewritten as 
\begin{equation}
R_{form}n_{H} = \frac{ N(H_{2}[v^{''}=0,J^{''}=4])  }{ N(HI) } 
\frac{A_{4 \rightarrow 2}}{(0.19 + 3.8p_{4,0})}
\end{equation}  
where $p_{4,0}$ is the radiative redistribution probability calculated by~\citet{jura75a}, $A_{J^{''} \rightarrow J^{'}}$ is the radiative transition probabilities for the mid-IR rotational $\Delta$$J$~=~4~$\rightarrow$~2 emission line~\citep{wolniewicz98},  
and N(HI) is the interstellar neutral hydrogen column density taken from direct Ly$\alpha$ measurements~\citep{diplas94}.  A similar equation can be constructed for $J^{''}$~=~5.   Under the assumption that the $J^{''}$~=~4 and 5 levels are predominantly populated by a combination of grain formation and radiative pumping (an assumption that is in question, \S1), we find that the average 
$R_{form}$$n_{H}$ for $\delta$~Sco and $\zeta$~Oph are 1.7~$\times$~10$^{-15}$ s$^{-1}$ and 2.3~$\times$~10$^{-13}$ s$^{-1}$, respectively.  

The product of the H$_{2}$ formation rate and the cloud density can be used to calculate the total photoabsorption rate ($\beta$) into the Lyman and Werner bands of H$_{2}$, which sets the balance for the excitation and dissociation of molecules in these clouds.  We find $\beta$($J^{''}$=0) = 4.9~$\times$~10$^{-13}$ s$^{-1}$ and 4.1~$\times$~10$^{-12}$ s$^{-1}$ for $\delta$~Sco and $\zeta$~Oph.  These values are 2~--~3 orders of magnitude lower than the canonical H$_{2}$ photoabsorption rate in the diffuse ISM, $\beta_{o}$~$\approx$~5~$\times$~10$^{-10}$ s$^{-1}$~\citep{jura74}, demonstrating that both of these clouds are heavily self-shielded, as expected given the damped $J^{''}$~=~0 and 1 absorption profiles for these sightlines.

Given the typical interstellar H$_{2}$ formation rate, $R_{form}$~$\approx$~3~$\times$~10$^{-17}$ cm$^{3}$ s$^{-1}$~\citep{jura75a,gry02}, the average H$_{2}$ cloud density on the $\delta$ Sco sightline is $n_{H}$~=~56 cm$^{-3}$.   As noted by~\citet{snow83}, the molecular formation rate in the $\rho$ Oph cloud on the $\delta$ Sco sightline may be lower than the typical ISM by factors of 2~--~3, therefore the $\delta$ Sco cloud density may be as high as 100~--~150 cm$^{-3}$ if this is the case.   

The density in the highly UV-irradiated portion of the $\zeta$ Oph sightline is $n_{H}$~$\approx$~7600 cm$^{-3}$, roughly consistent with previous estimates of the exterior region of the $\zeta$~Oph molecular absorber based on other spectral diagnostics~\citep{black73,morton75,wright79}, and higher than the bulk of the cooler molecular material on the sightline.  The combination of high density and high photoabsorption rate ($\beta_{\zeta Oph}$~$\sim$~10~$\times$~$\beta_{\delta Sco}$) suggests that a compressed portion of the interstellar cloud lies in close proximity to a strong source of far-UV irradiation (presumably $\zeta$ Oph itself).   The increase in N($J^{''}$~=~2~--~6) suggests spatial structure in the interface region on the scale of $\sim$130~AU (1.2\arcsec\ at the distance of $\zeta$~Oph, the angular displacement in the 40 years since the $Copernicus$ measurements). 
 Evidence for the $\zeta$~Oph interaction scenario has been clearly demonstrated by $WISE$ observations of a mid-IR bright bowshock in the direction of the space velocity of $\zeta$~Oph~\citep{peri12}.

The situation is reminiscent of the mid-IR bowshock observed around the runaway O9 V star HD~34078~\citep{france07}.  $FUSE$ observations of HD~34078 have shown the presence of vibrationally excited H$_{2}$ absorption arising from the compressed and strongly irradiated material swept up in the interaction between the star and the ambient ISM ($n_{H}$~$\sim$~10$^{4}$ cm$^{-3}$; Boiss{\'e} et al. 2005).~\nocite{boisse05}  $\zeta$~Oph is one of only a handful of other stars known to display vibrationally excited H$_{2}$ in its spectrum~\citep{federman95}, lending further support to the interaction scenario.  We predict that future high-resolution ($R$~$>$~10$^{5}$) far-UV (1000~--~1600~\AA) spectroscopy of the $\zeta$~Oph sightline will be able to isolate the velocity signature of this high-density, high-excitation molecular component.   


\acknowledgments
We acknowledge the hard work and dedication of the NASA Wallops~Flight~Facility/NSROC payload team, the Physical Sciences Laboratory at New Mexico State University, and the Navy team at WSMR that supported the 36.271 mission.  We are indebted to Ted Schultz for assistance with the design and fabrication of the SLICE electronics package.  KF acknowledges support through a NASA Nancy Grace Roman Fellowship during a portion of this work.   This work was further supported by NASA grants NNX10AC66G and NNX13AF55G to the University of Colorado at Boulder.  




\begin{deluxetable*}{c|cccccc}
\tabletypesize{\normalsize}
\tablecaption{SLICE Targets and Results. \label{lya_lines}}
\tablewidth{0pt}
\tablehead{
\colhead{Target} & \colhead{Sp Type}  &  \colhead{$d$} & \colhead{$T_{exp}$}  &
\colhead{log$_{10}$ N(HI)\tablenotemark{a} } &  \colhead{log$_{10}$ N(H$_{2}$)}  &  \colhead{$T$(H$_{2}$)}  \\
\colhead{} & \colhead{}  &  \colhead{(pc)} &
\colhead{(s)} &  \colhead{(cm$^{-2}$)}  &  \colhead{(cm$^{-2}$)}  &  
\colhead{(K)} 
}
\startdata
$\eta$ Uma & B3 V & 32  & 45  &  20.51~$\pm$~0.11 & $<$~14.86    & $\cdots$    \\
$\alpha$ Vir & B1 III & 77  & 30  &  $<$~19.66 & $<$~14.67    & $\cdots$    \\
$\delta$ Sco & B0 IV & 151  & 45  &  21.04~$\pm$~0.08      & 19.17~$\pm$~0.06    & $T_{01}$~=~67~$\pm$~1    \\
  &   &   &    &         &                                                                                           & $T_{exc}$~=~529~$\pm$~99  \\ 
$\zeta$ Oph & O9 V & 112  & 120  &  20.69~$\pm$~0.10  & 20.50~$\pm$~0.12   & $T_{01}$~=~66~$\pm$~3    \\
  &   &   &    &         &                                                                                           & $T_{exc}$~=~350~$\pm$~75  \\ 
\tableline
\tableline
$\delta$ Sco\tablenotemark{b} &   &    &    &        & N($J^{''}$=0)~=~18.93~$\pm$~0.01    &     \\
&   &    &    &                            & N($J^{''}$=1)~=~18.79~$\pm$~0.01    &     \\
&   &    &    &                            & N($J^{''}$=2)~=~16.09~$\pm$~0.05    &     \\
&   &    &    &                            & N($J^{''}$=3)~=~17.20~$\pm$~0.06    &     \\
&   &    &    &                            & N($J^{''}$=4)~=~14.65~$\pm$~0.10    &     \\
&   &    &    &                            & N($J^{''}$=5)~=~14.25~$\pm$~0.14    &     \\
&   &    &    &                            & N($J^{''}$=6)~=~13.33~$\pm$~1.41    &     \\
&   &    &    &                            & N($J^{''}$=7)~$<$14.51    &     \\
$\zeta$ Oph\tablenotemark{b} &   &    &    &        & N($J^{''}$=0)~=~20.26~$\pm$~0.03    &     \\
&   &    &    &                            & N($J^{''}$=1)~=~20.10~$\pm$~0.02    &     \\
&   &    &    &                            & N($J^{''}$=2)~=~18.88~$\pm$~0.04    &     \\
&   &    &    &                            & N($J^{''}$=3)~=~17.47~$\pm$~0.12    &     \\
&   &    &    &                            & N($J^{''}$=4)~=~16.82~$\pm$~0.62    &     \\
&   &    &    &                            & N($J^{''}$=5)~=~15.69~$\pm$~0.25    &     \\
&   &    &    &                            & N($J^{''}$=6)~=~14.08~$\pm$~0.27    &     \\
\enddata
\tablenotetext{a}{Derived from Ly$\alpha$ measurements~\citep{diplas94}, stellar Ly$\alpha$ contaminates N(HI) measurement for $\eta$~Uma and $\alpha$~Vir. } 
\tablenotetext{b}{Errors on individual H$_{2}$ rotational levels 
do not take into account continuum placement uncertainty (see text). } 
\end{deluxetable*}

\end{document}